\documentclass[nobibnotes,preprintnumbers,aps,prl,superscriptaddress,showpacs,twoside,sort&compress]{revtex4}				 %
\usepackage{graphicx}																												 %
\usepackage{titlesec}	
\usepackage{amsmath,amssymb}																										     %
\usepackage{fancyheadings}		
\usepackage{bm}																					 				 %
\usepackage{color}																								 %
\allowdisplaybreaks
\begin{document}																													 %
\vspace*{-0.3cm}																													 %

\title{Renormalization group calculation of dynamic exponent in the models E and F with hydrodynamic fluctuations.}

\author{M.~Dan\v{c}o}
\thanks{corresponding author; e-mail: danco@saske.sk}
\affiliation{Institute of Experimental Physics SAS, Watsonova 47, 040 01 Ko\v{s}ice, Slovakia}
\affiliation{BLTP, Joint Institute for Nuclear Research, Dubna, Russia}
\author{M.~Hnati\v{c}}
\affiliation{Institute of Experimental Physics SAS, Watsonova 47, 040 01 Ko\v{s}ice, Slovakia}
\affiliation{BLTP, Joint Institute for Nuclear Research, Dubna, Russia}
\affiliation{Institute of Physics, Faculty of Sciences, P. J. Safarik University, Park Angelinum 9, 041 54 Ko\v{s}ice, Slovakia}
\author{M.~V.~Komarova}
\affiliation{Department of Theoretical Physics, St. Petersburg University, Ulyanovskaya 1, St. Petersburg, Petrodvorets, 198504 Russia}
\author{T.~Lu\v{c}ivjansk\'y}
\affiliation{Institute of Physics, Faculty of Sciences, P. J. Safarik University, Park Angelinum 9, 041 54 Ko\v{s}ice, Slovakia}
\affiliation{Fakult{\"a}t f{\"u}r Physik, Universit{\"a}t Duisburg-Essen, D-47048 Duisburg, Germany}
\author{M.~Yu.~Nalimov}
\affiliation{Department of Theoretical Physics, St. Petersburg University, Ulyanovskaya 1, St. Petersburg, Petrodvorets, 198504 Russia}

\begin{abstract}
The renormalization group method is applied in order to analyze models E and F of critical dynamics in 
the presence of velocity fluctuations generated by
the stochastic Navier-Stokes equation. Results are given to the one-loop
approximation for the anomalous dimension $\gamma_{\lambda}$ and fixed-points' structure. 
The dynamic exponent $z$ is
calculated in the turbulent regime and stability of the fixed points for the standard model E is discussed.
\end{abstract}


\maketitle

%
\section{Introduction}
The liquid-vapor critical point, $\lambda$ transition in three-dimensional superfluid helium$\mbox{ }^4$He belong 
to famous examples of continuous phase transitions. It is known that regarding large-scale behavior such model lies
in the same universality class as the XY model \cite{hohenberg}. This fact is related to the observation that in 
both cases we have two-component order parameter. The divergent length at criticality reveals itself not only in
static but also dynamic properties of the system \cite{Tauber2014}. An analysis of infrared (IR) divergences and quantitative
predictions of universal quantities are therefore indispensable in understanding the dynamic behavior of spin
systems. When a system approaches the critical temperature at which a phase transition occurs, the
relaxation time $\tau$ diverges with the correlation length $\xi$ as $\xi^z$. The exponent $z$ defines 
 so-called dynamical critical exponent \cite{Tauber2014}. 
 
It is well known \cite{hohenberg,Tauber2014} that to a given static universality class different dynamic classes can be
assigned. In contrast to to the static models now the proper slow modes have to be identified and their governing
equation of motions specified. According to the general scheme  \cite{hohenberg,folk} the universal behavior in 
critical region of a superfluid is captured by model F. Recently this has been confirmed from microscopic 
principles \cite{hnatic} using the local interaction approximation. Model F reduces to model E in a limiting case 
in which coupling constants $g_2$ and $b$ (in our notation) are set to zero. Nowadays, there is no general consensus
which dynamic model (E or F) is genuine from the point of view of experimentally measurable quantities. In the 
corresponding static model, one of the $\omega$ indices coincides directly with the well-known, experimentally
measurable index $\alpha$ \cite{vasiliev}. The index $\alpha$ has been also determined in the framework of the RG 
approach using a resummation procedure \cite{kleinert} up to the four-loop perturbation precision 
and was measured in the Shuttle experiment \cite{lipa}. A contemporary accepted value is $\alpha = -0.0127$. The
negative sign of the index $\alpha$ ensures that $g_2^{*} = 0$ at the stable fixed point. This means that the stability
of model E can be considered as a particular 
realization of model F. Our main aim here is the calculation of the dynamic exponent $z$ and a stability analysis of
different IR scaling regimes due to an inclusion of velocity fluctuations in three-dimensional universality class of the XY model.
\section{Dynamics of Model F with Hydrodynamic Modes Activated}
Models E and F with an activated hydrodynamic modes were proposed and investigated by the RG method in \cite{hnatic,danco2}. Let us refer to model F with activated hydrodynamic modes as model $\text{F}_\text{h}$. Corresponding field theoretic model given in terms of the De Dominicis-Janssen action  \cite{janssen, dominicis} can be written as a sum 
$\mathcal{S}=\mathcal{S}_{\text{nc}}+\mathcal{S}_{\text{c}}+\mathcal{S}_{\text{v}}$, where
\begin{align}
  \mathcal{S}_{\text{nc}} & = 2\lambda_0 \psi^{\dagger'} \psi^{'} + \psi^{\dagger'}\Bigl\{-\partial_t \psi - \partial_i(v_i\psi) 
	+ \lambda_0(1+ib_0)\nonumber\\
	&\times[\partial^2\psi  - g_{01}(\psi^{\dagger}\psi)\psi/3 + g_{02}m\psi] 
	+i\lambda_0\psi[g_{07}\psi^{\dagger}\psi \nonumber\\
	&- g_{03}m +  g_{03}h_0 ]   \Bigr\} + \mbox{H.c}
	\label{eq:action_psi}
\end{align}
 describes dynamics of the nonconserved order parameter fields $(\psi,\psi^\dagger)$ and H.c. stands for a 
 hermitian conjugate. Further, the action
\begin{align}
  \mathcal{S}_{\text{c}} & =  - \lambda_0 u_0 m' \partial^2 m' + m'\Bigl\{ -\partial_t - 
	v_i\partial_i m - \lambda_0 u_0 \partial^2[-m \nonumber\\
	& + g_{06}\psi^{\dagger}\psi + h_0]
	+ i\lambda_0 g_{05}[\psi^{\dagger}\partial^2\psi-\psi\partial^2\psi^{\dagger}] \Bigr\} 
	\label{eq:action_m}
\end{align}
describes dynamics of the conserved field $m$, which can be interpreted as a linear combination of density and temperature field \cite{vasiliev}. The velocity fluctuations are governed by the following action
\begin{align}
  \mathcal{S}_{\text{v}} =  \frac{1}{2}v'_iD^v_{ij}{v'_j} + { v'_i} 
  \{ -\partial_t {v_i} + \nu_0 \partial^2 {v_i} - v_j\partial_j v_i \},
  \label{eq:action_vel}
\end{align}
where the explicit form of $D^v_{ij}$ can be found, e.g., in \cite{vasiliev} or \cite{danco2}. For simplicity integrals over spatial and time variables in (\ref{eq:action_psi}-\ref{eq:action_vel}) have been omitted and prime fields denote Martin-Siggia-Rose response fields \cite{MSR}. 
 The intermode coupling of fields $\psi$ and $\psi^\dagger$ with the field $m$ in (\ref{eq:action_psi}) and (\ref{eq:action_m}) corresponds physically to the exact relation between the phase of the complex order parameter and the chemical potential \cite{hohenberg}. The interactions with velocity field $v_i$  are introduced \cite{vasiliev} via standard replacement $\partial_t\rightarrow\partial_t + v_i\partial_i$ and from this point of view the passive advection is introduced into the model. We consider the velocity field to
be incompressible which is tantamount to the condition $\partial_iv_i =0$ \cite{vasiliev}.

The field-theoretic formulation summarized in Eqs. (\ref{eq:action_psi})-(\ref{eq:action_vel}) has an advantage to be amenable to the full machinery of field
theory \cite{vasiliev}. Near criticality large fluctuations on all scales dominate the behavior of the system, which results into the infrared 
divergences in Feynman diagrams. The RG technique allows us to deal with them and as a result provides us with information about the scaling behavior of the system. Moreveover, it also leads to a pertubative computation of critical exponent in a formal expansion around the upper critical dimension. In contrast to the standard $\varphi^4$-theory we have to deal with two-parameter expansion $(\varepsilon,\delta)$, where $\varepsilon$ is the deviation from the upper critical dimension $d_c=4$, and $\delta$ is describes nonlocal behavior of random noise for velocity fluctuations. It follows the approach suggested in original work \cite{honnal}. The model under consideration is augmented \cite{vasiliev,danco} by the following relations between the charges 
\begin{equation}
  g_{05} = g_{03}, \quad g_{06} = g_{02}, \quad g_{07}=g_{02}g_{03}.
  \label{eq:charges}
\end{equation}
The introduction of the new coupling constants are needed in order to restore the multiplicative renormalizability of the model.
 Details of the perturbative renormalization grop calculations can be found in \cite{danco}. Here, we concentrate on the 
 calculation of the dynamic exponent $z$, which has not been done previously. To this end we need to determine the
 anomalous dimension $\gamma_{\lambda}$ \cite{vasiliev,Tauber2014}, because the latter determines $z$ through the relation
\begin{equation}
  z = 2 - \gamma_{\lambda}^{*},
  \label{eq:dyn_scaling}
\end{equation}
where the asterisk denotes the fixed point's value. The one-loop expression for  $\gamma_{\lambda}$ can be written in the form
\begin{align}
  \gamma_{\lambda} & =  \frac{1}{\left[b^2+(1+u)^2\right]^3}\Bigl(g_2^2 A_1 + g_2 g_3 A_2 + g_3^2 A_3  +  g_2g_5A_4 \nonumber \\  
  &+ g_3g_5A_5 + g_2g_6A_6 + g_3g_6A_7 + g_4A_8 \Bigr), 
  \label{eq:gamma}
\end{align}
where
\begin{align}
	A_1 & =  b^6(3u-1)-b^4[3+u(u^2-9u-3)] \nonumber \\ 
	    & +  3b^2(1+u)(2u^2-1)-(1+u)^3, \nonumber \\
	A_2 & =  -6b^5u + 2b^3u[u^2-6u-6]-6bu(1+u)^2, \nonumber \\
	A_3 & =  b^4(1+3u)-b^2(1+u)[u^2-4u-2] +  (1+u)^3, \nonumber \\
	A_4 & =  2ub^5 + b^3u[4+2u-3u^2]  -  bu(1+u)^2[u^2+2u-2],  \nonumber \\
	A_5 & =  -2b^4u + 3b^2u^2(1+u)+u(1+u)^3(2+u), \nonumber \\
	A_6 & =  -2ub^6 + b^2u(1+u)[u(u-2)(4+u)-6] \nonumber \\ 
	    & + ub^4(u-3)(2+3u) - u(1+u)^3(2+u), \nonumber \\
	A_7 & =  2ub^5+b^3u(4+2u-3u^2) \nonumber \\ 
	    & + bu(2 + 2u-3u^2-4u^3-u^4),  \nonumber \\
	A_8 & =  -\frac{3u_1^2(1+u_1)}{8}.
\end{align}
These relation are in agreement with \cite{danco,danco2} in the special limit obtained for $b=g_2=g_6=g_7=0$, which corresponds
to model $\text{E}_\text{h}$.

\section{Scaling Regimes and Fixed-Point Structure}
Scaling regimes are associated with fixed points of the RG equations. The fixed points are defined as the points $g^{*}=(g_1^*,\cdots,g_7^*,u^*,u^*_1)$  at which all $\beta$-functions simultaneously vanish, i.e.
\begin{equation}
  \beta_e (g^{*}) = 0, \quad e \in \{g_1,\cdots,g_7,u,u_1\}.
  \label{eq:beta}
\end{equation}
 The type of the fixed point is determined by the eigenvalues of the matrix of its first derivatives $\omega =\{ \omega_{ij} = \partial \beta_i / \partial e_j \}$, where $\beta_i$ is the full set of $\beta$ functions and $e_j$ is the full set of charges. The IR-asymptotic
behavior is governed by IR-stable fixed points, for which all real parts of eigenvalues of the matrix $\omega$ are positive. In fact, there are two physically possible and interested regimes. 
The first one is the regime with hydrodynamic fluctuations near the thermodynamic equilibrium that corresponds to the values $\varepsilon=1$ and $\delta=-1$. The second one is the Kolmogorov turbulent regime with $\varepsilon=1$ and $\delta=4$. 

\section{Model $\text{F}_\text{h}$}
A majority of the fixed points can be found only in a numerical fashion. A fraction of them can be immediately discarded due to  unacceptable values of physical parameters. This is why we have attempted to investigate the system specifically in different regimes, rather then
solving it directly. In the turbulent scaling regime numerical analysis reveals an IR stable fixed point with coordinates
\begin{align}
  &g_4^{*} =  10.\overline{6},\quad u^{*}=1,\quad  u_1^{*} = 0.7675919, \nonumber\\
  &b^{*}  =  g_1^{*} = g_2^{*} = g_3^{*} = g_5^{*} = g_6^{*} = g_7^{*} = 0,
  \label{eq:coord}
\end{align}
where the overline symbol stands for a repeating decimal. At this fixed point the anomalous dimension $\gamma_{\lambda}^{*}$ and
 eigenvalues of the $\omega$ matrix are
\begin{equation}
  \gamma_{\lambda}^{*} = 1.\overline{3},\qquad \omega = \{2.087, 1.666, 0.833, 4, 2.921 \}.
  \label{eq:anomal}
\end{equation} 
The critical dynamic exponent $z$ corresponding to this regimes is $z=0.\overline{6}$. It is worth to mention that the multi-loop calculations could change the stability of a given fixed point. In the thermal equilibrium the numerical analysis of model $\text{F}_\text{h}$ has not exhibited the existence of the IR stable fixed point. 
\section{Model $\text{E}_\text{h}$}
The fixed points of model $\text{E}_\text{h}$ turn out to be unstable in the context of model $\text{F}_\text{h}$, but this instability could 
be just a consequence of the used low approximation. Let us include the fixed points of model $\text{E}_\text{h}$ into consideration. The most interesting result is how to analyze the stability for the standard model E. Indeed, some of the fixed points of model $\text{E}_\text{h}$ correspond to the standard model E \cite{peliti}. They must obey the conditions $g_4=u_1=0$ and $g_3=g_5$. One of such points is a dynamical fixed point 
\begin{align}
  g_1^{*} & = \frac{3}{5}\varepsilon,  &g_3^{*}& = \varepsilon^{1/2}, &g_5^{*}& = \varepsilon^{1/2}, \nonumber \\ 
  u^{*} & =  1, &g_4^{*}&=0, &u_1^{*}&=0,
\end{align}
where $\gamma_{\lambda}^{*} = \varepsilon/2$ and hence the exponent $z$ is given by $z = d/2$. The second is called a weak-scaling point \cite{peliti}
\begin{align}
  g_1^{*} & =  \frac{3}{5}\varepsilon, &\left(\frac{g_3^2}{u}\right)^{*}& = \frac{2}{3}\varepsilon,  
  &\left(\frac{g_5^2}{u}\right)^{*}& = \frac{2}{3}\varepsilon, \nonumber \\
  \left(\frac{1}{u}\right)^{*} & =  0, &g_4^{*}& = 0, &u_1^{*}& = 0,
\end{align}
where the RG function $\gamma_{\lambda}^{*} = 2\varepsilon/3$ and then in this case the exponent $z$ is nontrivial $z = 2 - 2\varepsilon/3$. It is unknown which of these two points is stable for the standard model E. In the framework of model $F_h$ these two points have the following $\omega$ indices
\begin{align}
  &\omega_1 \in \{-0.1\varepsilon, 0, 0.055\varepsilon, 0.25\varepsilon,0.75\varepsilon,\varepsilon,
  1.5\varepsilon,1.92\varepsilon,-\delta \}, \nonumber \\
  &\omega_2 \in \{ -0.33\varepsilon, -0.01\varepsilon, -0.05\varepsilon, 0.66\varepsilon, \varepsilon,
  1.3\varepsilon, 2.15\varepsilon, -\delta \}, 
\end{align}
where the repeated $\omega$ indices are omitted. This means that the dynamical fixed point seems to be more IR stable with respect to hydrodynamics effects.

\section{Conclusions}
Models E and F of critical dynamics have been studied in the critical region with both critical and 
velocity fluctuations taken into account. The anomalous dimension $\gamma_{\lambda}$ has been computed 
for model F to the one-loop approximation and the fixed points' structure has been partly analyzed. For
model $\text{F}_\text{h}$ the IR stable fixed point has been found in the turbulent scaling regime, where
the dynamic critical exponent equals $z=0.\overline{6}$. The corollary gained from the analysis of the thermal 
equilibrium regime suggests that the one-loop calculation of models $\text{E}_\text{h}$ and $\text{F}_\text{h}$  
is not sufficient to make an ultimate conclusion about the stability of fixed points. The first question that has
to be addressed is which of these points is stable and corresponds to the physical reality. The available results
do not allow us to give a clear answer to this question, therefore, an urgent task is to consider higher order 
terms in the perturbation theory.
\section{Acknowledgement}
The work was supported by VEGA Grant No. 1/0222/13 of the Ministry of Education, Science, Research and Sport of the Slovak Republic. We would like to thank Dr. Martin Vala and the project Slovak Infrastructure for High Performance Computing (SIVVP) ITMS 26230120002.

\end{document}